\documentstyle[12pt,epsfig]{article}
\newcommand{\nd}{\noindent}
\newcommand{\beq}{\begin{equation}}
\newcommand{\eeq}{\end{equation}}
\newcommand{\barr}{\begin{eqnarray}}
\newcommand{\earr}{\end{eqnarray}}
\newcommand{\ba}{\begin{array}}
\newcommand{\ea}{\end{array}}
\newcommand{\bfp}{\mbox{\boldmath $p$}}
\newcommand{\bfk}{\mbox{\boldmath $k$}}
\newcommand{\bfy}{\mbox{\boldmath $y$}}

\newcommand{\ga}{\gamma}

\newcommand{\ep}{\varepsilon}
\newcommand{\la}{\lambda}

\newcommand{\pup}{p^\uparrow}
\newcommand{\pdown}{p^\downarrow}
\newcommand{\NP}[1]{{\it Nucl.\ Phys.}\ {\bf #1}}
\newcommand{\ZP}[1]{{\it Z.\ Phys.}\ {\bf #1}}
\newcommand{\PL}[1]{{\it Phys.\ Lett.}\ {\bf #1}}
\newcommand{\PR}[1]{{\it Phys.\ Rev.}\ {\bf #1}}
\newcommand{\PRL}[1]{{\it Phys.\ Rev.\ Lett.}\ {\bf #1}}

\newcommand{\SNP}[1]{{\it Sov.\ J.\ Nucl.\ Phys.}\ {\bf #1}}
\def\lsim{\mathrel{\rlap{\lower4pt\hbox{\hskip1pt$\sim$}}\raise1pt\hbox{$<$}}}
\def\gsim{\mathrel{\rlap{\lower4pt\hbox{\hskip1pt$\sim$}}\raise1pt\hbox{$>$}}}
\def\nostrocostruttino#1\over#2{\mathrel{\mathop{\kern 0pt \rlap
{\hbox{$#1$}}} \hbox{\kern-.135em $#2$}}}
\def\sumint{\nostrocostruttino \sum \over {\displaystyle\int}}
\begin{document}
\begin{flushright}
DFTT 48/94 \\
INFNCA-TH-94-27 \\
hep-ph/9503290 \\
Revised version - July 1995
\end{flushright}
\vskip 1.5cm
\begin{center}
{\large \bf
SINGLE SPIN ASYMMETRY FOR \mbox{\boldmath{$p^{\uparrow}p \to \pi X$}} \\
IN PERTURBATIVE QCD
}\\
\vskip 1.5cm
{\sf M.\ Anselmino$^1$, M.\ Boglione$^1$, F.\ Murgia$^2$}
\vskip 0.8cm
{$^1$ Dipartimento di Fisica Teorica, Universit\`a di Torino and \\
      INFN, Sezione di Torino, Via P. Giuria 1, 10125 Torino, Italy\\
\vskip 0.5cm
 $^2$ INFN, Sezione di Cagliari, Via A. Negri 18, 09127 Cagliari, Italy } \\
\end{center}
\vskip 1.5cm
\noindent
{\bf Abstract:} \\
Within the QCD-improved parton model and assuming the factorization theorem
to hold in the helicity basis and for higher twist contributions, we show
how non zero single spin asymmetries in hadron-hadron high energy and
moderately large $p_T$ inclusive processes can be obtained, even in massless
perturbative QCD, provided the quark intrinsic motion is taken into account.
A simple model is constructed which reproduces the main features of the data
on the single spin asymmetry observed in inclusive pion production in $p\,p$
collisions.
\newpage
\pagestyle{plain}
\setcounter{page}{1}
\nd
{\bf 1 Introduction}
\vskip 6pt

Single spin asymmetries in high energy and moderately large $p_T$ inclusive
hadronic processes have recently received much attention, both experimentally
\cite{ada1}-\cite{ada4} and theoretically \cite{rys}-\cite{efr2}.
Whereas they are expected to vanish at leading twist in massless perturbative
QCD, higher twist effects might still be important in the kinematical region
of the available data and may give origin to non zero values. Among the
attempted explanations, quark-gluon correlations \cite{rys,qiu,efr2} and
transverse $k_\perp$ effects in the quark distribution \cite{siv1,siv2} or
fragmentation functions \cite{art} have been considered.

We analyse here the large single spin asymmetries observed in the collision
of transversely polarized protons off unpolarized protons, with the production
of pions with $p_T$ values up to 2 GeV/$c$,
\beq
p^{\uparrow} p \rightarrow \pi X \,.
\label{pro}
\eeq
Several experimental results are available for such processes
\cite{ada2}-\cite{ada4}
and show clear patterns in the dependence of the spin asymmetry
\beq
A_N = \frac{ d\sigma^{\uparrow} - d\sigma^{\downarrow} }
           { d\sigma^{\uparrow} + d\sigma^{\downarrow} }
\label{asy}
\eeq
on $x_F = 2 p_L/\sqrt{s}$, where $p_L$ is the pion longitudinal momentum
in the $p\,p$ c.m. system, and $\sqrt{s}$ is the total c.m. energy. The proton
spin is perpendicular to the scattering plane and there exist data for
$\pi^{\pm}$ and $\pi^{0}$; proton spin orientations parallel to
the scattering plane would give, by parity invariance, zero single spin
asymmetries. A striking dependence of $A_N$ on $p_T$ which was reported
at $x_F \simeq 0$ \cite{ada3} is now believed, after a more careful analysis
of the data \cite{nur}, not to be observed anymore; $A_N(x_F \simeq 0)$ shows
no sign of dependence on $p_T$ and remains consistent with zero in the whole
range of $p_T$ explored so far, ($0 \lsim p_T \lsim 4$) GeV/$c$ \cite{nur}.

Some previous data \cite{ant}-\cite{sar} seem to be in disagreement with
the findings of Refs. \cite{ada1}-\cite{ada4}, \cite{nur}; in particular, they
do not show sizeable values of $A_N$ for $\pi^-$ \cite{bon,sar}, show strong
$p_T$ dependences for $\pi^0$ \cite{ant} and $\pi^+$ \cite{sar} and have
contradictory results on $x_F$ dependences \cite{bon,sar}. However, all
these data come from experiments at much smaller energy (proton beams of
13.3, 18.5 or 24 GeV/$c$) than those of Refs. \cite{ada1}-\cite{ada4},
\cite{nur} (200 GeV/$c$) and we do not consider them here because they
might be outside the range of applicability of perturbative QCD.

The usual QCD description of large $p_T$ inclusive production is based on the
factorization theorem, according to which the cross-section for the production,
say, of a large $p_T$ pion in the collision of two protons is given by the
convolution of an elementary cross-section with the number densities of quarks
and gluons inside the protons (distribution function) and the number density
of pions inside a quark or gluon (fragmentation function). The elementary
cross-section contains all the dynamical and quantum-mechanical
information on the constituent interaction, whereas the distribution and
fragmentation functions are phenomenological ways of modeling the
non-perturbative long distance physics: they can be measured in other processes
and their large $Q^{2}$ evolution is given by perturbative QCD.

We know that the above hard scattering scheme works well for unpolarized
processes and indeed it has been tested in many experiments. It has also
been generalized to the polarized case \cite{col2}, so that it may be
applied to the description of several processes involving polarized hadrons
\cite{col3}; however, the existing spin data do not allow yet a definite test
of its validity.

In Section 2 we adapt the formalism of Refs. \cite{col} and \cite{col2}
to the case of the single spin asymmetry (\ref{asy}). In order not to obtain a
zero result higher twist effects have to be introduced; this can be done
at different stages and several suggestions or attempts have been proposed in
the literature \cite{siv1}-\cite{art}, \cite{szw,efr}.
Single spin effects in the elementary reactions alone are bound
to be proportional to $\alpha_s m_q/\sqrt{s}$ \cite{kan}, where $m_q$ is the
quark mass, and are then expected to be negligible at high energies, even
taking constituent quark masses into account \cite{siv1,szw,efr}.
Spin effects might then be present in the distribution or fragmentation
functions: the former has been suggested by Sivers \cite{siv1,siv2} and
the latter by Collins \cite{col}, whose idea has been further
developed and applied in Ref. \cite{art}. Qiu and Sterman \cite{qiu} have used
both higher order elementary interactions and higher twist distribution
functions to predict a sizeable single spin asymmetry in large $p_T$ direct
photon inclusive production, $\pup p \to \gamma X$.

The approach we describe here is equivalent, although derived in a different
way, to the suggestion of Refs. \cite{siv1,siv2} and supports it; we then
discuss (Section 3) a simple model which implements the idea and gives very
good agreement with the data. The approach of Refs. \cite{siv1,siv2} has been
criticized in Ref. \cite{col} on the ground of violating the time reversal
invariance of QCD. This is true only at leading twist order, if soft initial
state interactions between the colliding protons are neglected, which need not
be the case in Sivers or our model; we will further comment on this in
Section 2. A short conclusion is given in Section 4.

\vskip 12pt
\nd
{\bf 2 The single spin asymmetry in the hard scattering scheme}
\vskip 6pt

Let us then consider the process (1), supposing the initial protons moving
along the $z$-axis and choosing $xz$ as the production plane; the incoming
proton is polarized parallel ($\uparrow$) or opposite ($\downarrow$) the
$\hat{\bfy}$-direction so that, in the helicity basis,
\beq
|\uparrow \, \rangle = \frac{1}{\sqrt 2} (|+\rangle +i |-\rangle)
\label{up}
\eeq
\beq
|\downarrow \, \rangle = \frac{-1}{\sqrt 2} (|+\rangle -i |-\rangle) \,.
\label{down}
\eeq

According to the QCD factorization theorem the differential cross-section
for the hard scattering of a polarized proton with spin $\uparrow$
(and similarly for spin $\downarrow$) on an unpolarized target proton,
resulting in the inclusive
production of a pion with energy $E_\pi$ and three-momentum $\bfp_\pi$,
$\pup p \to \pi X$, can be written as \cite{col,col2,col3}
\barr
\frac{E_\pi \, d\sigma^{\pup p \to \pi X}} {d^{3} \bfp_\pi} &\sim &
{1\over 2} \sum_{a,b,c,d} \>
\sum_{\la^{\,}_a, \la^{\prime}_a; \la^{\,}_b; \la^{\,}_c, \la^{\prime}_c;
\la^{\,}_d}
\int dx_a \, dx_b \, {1 \over z} \label{gennok} \\
& & \rho_{\la^{\,}_a, \la^{\prime}_a}^{a/\pup} \,
f_{a/\pup}(x_a) \,f_{b/p}(x_b) \,
\hat M_{\la^{\,}_c, \la^{\,}_d; \la^{\,}_a, \la^{\,}_b} \,
\hat M^*_{\la^{\prime}_c, \la^{\,}_d; \la^{\prime}_a, \la^{\,}_b} \,
D_{\pi/c}^{\la^{\,}_c,\la^{\prime}_c}(z) \,,
\nonumber
\earr
where $f_{a/\pup}(x_a)$ is the number density of partons $a$ with momentum
fraction $x_a$ inside the polarized proton [similarly for $f_{b/p}(x_b)$]
and $\rho_{\la^{\,}_a, \la^{\prime}_a}^{a/\pup}(x_a)$ is the helicity density
matrix of parton $a$ inside the polarized proton $\pup$.
The $\hat M_{\la^{\,}_c, \la^{\,}_d; \la^{\,}_a, \la^{\,}_b}$'s
are the helicity amplitudes for the elementary process $ab \to cd$;
if one wishes to consider higher order (in $\alpha_s$) contributions also
elementary processes involving more partons should be included.
$D_{\pi/c}^{\la^{\,}_c,\la^{\prime}_c}(z)$ is the product of
{\it fragmentation amplitudes}
\beq
D_{\pi/c}^{\la^{\,}_c,\la^{\prime}_c} = \sumint_{X, \la_{X}}
{\cal D}_{\la^{\,}_{X}; \la^{\,}_c} \,
{\cal D}^*_{\la^{\,}_{X}; \la^{\prime}_c}
\label{framp}
\eeq
where the $\sumint_{X, \la_{X}}$ stays for a spin sum and phase space
integration of the undetected particles, considered as a system $X$.
The usual unpolarized fragmentation function $D_{\pi/c}(z)$, {\it i.e}
the density number of pions resulting from the fragmentation of
an unpolarized parton $c$ and carrying a fraction $z$ of its momentum
is given by
\beq
D_{\pi/c}(z) = {1\over 2} \sum_{\la^{\,}_c}
D_{\pi/c}^{\la^{\,}_c,\la^{\,}_c}(z)\,.
\label{fr}
\eeq
For simplicity of notations we have not indicated in Eq. (\ref{gennok})
the $Q^2$ scale dependences in $f$ and $D$; the variable $z$ is related to
$x_a$ and $x_b$ by the usual imposition of energy momentum conservation
in the elementary 2 $\to$ 2 process \cite{bla}; we have skipped, for the
moment, some spin independent kinematical factors\footnote{Equation
(\ref{gennok}) holds as an equality if the elementary amplitudes are
normalized so that
$(1/4) \sum_{\la^{\,}_a, \la^{\,}_b, \la^{\,}_c, \la^{\,}_d}
|\hat M_{\la^{\,}_a, \la^{\,}_b, \la^{\,}_c, \la^{\,}_d}|^2 = (1/\pi)
(d\hat\sigma / d\hat t)$}, but we explicitely kept
the factor 1/2 to remind that an average has been taken over the
helicities of the unpolarized parton $b$ (quark or gluon).

Eq. (\ref{gennok}) holds at leading twist and large $p_T$ values of the
produced pion; the intrinsic $\bfk_\perp$ of the partons have
been integrated over and collinear configurations dominate both the
distribution functions and the fragmentation processes; one can then see
that, in this case, there cannot be any single spin asymmetry. In fact,
total angular momentum conservation in the (forward) fragmentation process
[see Eq. (\ref{framp})] implies $\la^{\,}_c = \la^{\prime}_c$; this, in turns,
together with helicity conservation in the elementary processes, implies
$\la^{\,}_a = \la^{\prime}_a$. If we further notice that, by parity invariance,
$D_{\pi/c}^{\la^{\,}_c,\la^{\,}_c}$ does not depend on $\la_c$ and that
$\sum_{\la^{\,}_b, \la^{\,}_c, \la^{\,}_d}
|\hat M_{\la^{\,}_c, \la^{\,}_d; \la^{\,}_a, \la^{\,}_b} |^2$ is
independent of $\la^{\,}_a$ we remain with
$\sum_{\la^{\,}_a} \rho_{\la^{\,}_a, \la^{\,}_a}^{a/\pup} = 1$. Moreover,
in the absence of intrinsic $\bfk_\perp$ and initial state interactions,
the parton density numbers $f_{a/\pup}(x_a)$ cannot depend on the proton
spin and any spin dependence disappears from Eq. (\ref{gennok}), so that
\beq
d\sigma^{\pup p \to \pi X} - d\sigma^{\pdown p \to \pi X} = 0 \,.
\eeq

Eq. (\ref{gennok}) can be generalized with the inclusion of intrinsic
$\bfk_\perp$ \cite{col} and this can avoid the above conclusion; for example
\cite{col}, the observation of a non zero $\bfk_\perp$ of a final particle
$C$ with respect to the axis of the jet generated by parton $c$ does not imply
any more $\la^{\,}_c = \la^{\prime}_c$ and allows a non zero value of the
asymmetry
\beq
d\sigma^{\pup p \to \pi,\bfk_\perp \, X} -
d\sigma^{\pdown p \to \pi,\bfk_\perp \, X} \,.
\label{cef}
\eeq

The above asymmetry (\ref{cef}) is related to the so called Collins
\cite{art, kot, tan} or sheared jet \cite{car} effect;
it requires the measurement of the azimuthal angle $\phi$ of the outgoing
hadron around the jet axis, but, apart from a small $\sin \phi$ dependence,
it is a leading twist effect and it depends on some non perturbative quark
fragmentation analysing power. When integrating over the azimuthal angle
the effect might not entirely disappear because of some $\phi$ dependence
in the elementary parton interaction. This idea was exploited in
Ref. \cite{art} where, essentially, the parton $c$ is produced in the
forward direction and the final hadron $p_T$ is due to its transverse
$k_\perp$ inside the jet. One cannot expect such a model to work at large
$p_T$.

Another possible $\bfk_\perp$ effect, suggested by Sivers \cite{siv1, siv2},
may originate in the distribution functions. To see how this comes out
from the general scheme we rewrite Eq. (\ref{gennok}) taking into account
the parton intrinsic momentum in the number density $f_{a/p}$:
\barr
\frac{E_\pi \, d\sigma^{\pup p \to \pi X}} {d^{3} \bfp_\pi}&\sim&
{1\over 2} \sum_{a,b,c,d} \>
\sum_{\la^{\,}_a, \la^{\prime}_a; \la^{\,}_b; \la^{\,}_c, \la^{\prime}_c;
\la^{\,}_d}
\int d^2\bfk_{\perp a} dx_a \, dx_b \, {1 \over z} \label{genk} \\
&& \rho_{\la^{\,}_a, \la^{\prime}_a}^{a/\pup} \,
\hat f_{a/\pup}(x_a, \bfk_{\perp a}) \,f_{b/p}(x_b) \,
\hat M_{\la^{\,}_c, \la^{\,}_d; \la^{\,}_a, \la^{\,}_b} \,
\hat M^*_{\la^{\prime}_c, \la^{\,}_d; \la^{\prime}_a, \la^{\,}_b} \,
D_{\pi/c}^{\la^{\,}_c,\la^{\prime}_c}(z) \,,
\nonumber
\earr
where $\hat f$ denotes the $\bfk_\perp$ dependent number density.

We can now argue, as in the previous case when no $\bfk_\perp$ was taken
into account, that angular momentum, helicity and parity conservation
eliminate all dependences on the parton helicities in Eq. (\ref{genk});
however, a dependence on the hadron spin may remain in
$\hat f_{a/\pup}(x_a, \bfk_{\perp a})$, analogously to the Collins effect
in the fragmentation process. Then one has \cite{siv1, siv2}:
\barr
&& \frac{E_\pi \, d\sigma^{\pup p \to \pi X}} {d^{3} \bfp_\pi} -
\frac{E_\pi \, d\sigma^{\pdown p \to \pi X}} {d^{3} \bfp_\pi} \sim \nonumber\\
&& {1\over 4} \sum_{a,b,c,d} \>
\sum_{\la^{\,}_a, \la^{\,}_b; \la^{\,}_c, \la^{\,}_d}
\int d^2\bfk_{\perp a} \, dx_a \, dx_b \, {1 \over z}
\label{gendif} \\
&& \times \left[ \hat f_{a/\pup}(x_a, \bfk_{\perp a})
    - \hat f_{a/\pdown}(x_a, \bfk_{\perp a}) \right] \, f_{b/p}(x_b) \,
\left| \hat M_{\la^{\,}_c, \la^{\,}_d; \la^{\,}_a, \la^{\,}_b} \right|^2
D_{\pi/c}(z) \,. \nonumber
\earr

Several comments are now in order.

There is a new quantity which appears in Eq. (\ref{gendif}):
\barr
\Delta^Nf_{a/\pup}(x_a,\bfk_{\perp a}) &\equiv& \sum_{\la^{\,}_a} \left[
\hat f_{a, \la^{\,}_a / \pup} (x_a,\bfk_{\perp a})
- \hat f_{a, \la^{\,}_a / \pdown} (x_a,\bfk_{\perp a}) \right]
\label{isiv1}\\
&=& \sum_{\la^{\,}_a}
\left[ \hat f_{a, \la^{\,}_a / \pup} (x_a,\bfk_{\perp a})
- \hat f_{a, \la^{\,}_a / \pup} (x_a, -\bfk_{\perp a}) \right] \,,
\label{isiv2}
\earr
where $\hat f_{a, \la^{\,}_a / p^{\uparrow (\downarrow)}}(x_a,\bfk_{\perp a})$
is the number density of partons $a$ with helicity $\la^{\,}_a$,
momentum fraction $x_a$ and intrinsic transverse momentum $\bfk_{\perp a}$
in a transversely polarized proton [with spin $\uparrow$ or $\downarrow$
according to Eq. (\ref{up}) or (\ref{down})]. Eq. (\ref{isiv2}) follows from
Eq. (\ref{isiv1}) by rotational invariance and explicitely shows that
$\Delta^Nf_{a/\pup}(x,\bfk_{\perp}) = 0$ when $\bfk_{\perp}=0$.

This new quantity can be regarded as a single spin asymmetry or analysing
power for the $\pup \to a + X$ process; if we define the polarized number
densities in terms of {\it distribution amplitudes} as
\beq
\hat f_{a, \la^{\,}_a/ \pup}(x_a, \bfk_{\perp a}) = \sumint_{X_p,\la_{X_p}}
|{\cal G}^{a/p}_{\la_{X_p}, \la^{\,}_a; \uparrow}(x_a,\bfk_{\perp a})|^2
\label{g}
\eeq
then we have, in the helicity basis,
\barr
\Delta^Nf_{a/\pup}(x_a,\bfk_{\perp a}) &=&
\sumint_{X_p,\la_{X_p}} \sum_{\la^{\,}_a} 2\,\mbox{Im} \, \left[
{\cal G}^{a/p}_{\la_{X_p}, \la^{\,}_a;+}(x_a,\bfk_{\perp a}) \,\,
{\cal G}^{a/p\,*}_{\la_{X_p},\la^{\,}_a;-}(x_a,\bfk_{\perp a}) \right]
\nonumber \\
&\equiv& 2 \, I^{a/p}_{+-}(x_a,\bfk_{\perp a}) \,.
\label{iap}
\earr

Eq. (\ref{iap}) simply follows from Eqs. (\ref{isiv1}) and (\ref{g})
via Eqs. (\ref{up}) and (\ref{down}) and shows the non diagonal nature,
in the helicity indices, of $I^{a/p}_{+-}(x_a,\bfk_{\perp a})$.

Collins \cite{col} has argued that a non zero value of
$I^{a/p}_{+-}(x,\bfk_{\perp})$ is forbidden by the time reversal invariance
of QCD; his argument is based on the analysis of the non diagonal matrix
elements of the leading twist quark operator $\bar\psi \ga^+ \psi$, whose
diagonal matrix elements are related to the distribution functions
$f_{q/p}(x, \bfk_\perp)$. For such operator his argument is correct and indeed
time reversal invariance forces the matrix elements between proton states
with different helicities to be zero. In a more physical language this
amounts to say that single spin asymmetries for the process
$\pup \to qX$ are forbidden by parity and time reversal invariance,
which is true. However, as we said, initial state interactions (like soft
gluon exchanges) between the incoming protons must certainly occur, and,
at least in cases in which their neglect gives a zero result, one should
consider them and relate the distribution functions to the inclusive
cross-section for the process $p_1 p_2 \to qX$; single (transverse)
spin asymmetries are then certainly allowed (as the problem
we are studying here confirms) via time reversal invariant scalar
quantities like $\ep_{\mu\nu\rho\sigma}p_1^\mu p_2^\nu q^\rho s_1^\sigma$.

These soft gluon and initial state interactions which correlate partons from
different hadrons and allow a non zero value of $I^{a/p}_{+-}$ can only
survive at higher twist; it has been shown \cite{col2} that at leading
twist-2 the proof of the factorization theorem for the unpolarized case
-- with the cancellation of all soft gluon contributions -- holds for
the polarized case as well. Eq. (\ref{gendif}), as it will be shown in the
next Section, only gives higher twist contributions; assuming a non zero
value of $I^{a/p}_{+-}$ in the factorized structure of Eq. (\ref{gendif})
amounts to assume the validity of the factorization theorem beyond leading
twist.

In the operator language this approach has been advocated by Qiu and Sterman
\cite{qiu} who use generalized factorization theorems valid at higher twist
and relate non zero single spin asymmetries in $p\,p$ collisions to the
expectation value of a higher twist operator, a twist-3 parton distribution,
which explicitly involves correlations between the two protons and combines
quark fields with a gluonic field strength. However, they still consider
only collinear partonic configurations so that, in order to obtain non
zero results, they have to take into account the contributions of higher
order elementary interactions.

Our function $I^{a/p}_{+-}(x_a,\bfk_{\perp a})$ introduced in Eqs. (\ref{iap})
and (\ref{isiv1}) can be considered as a new phenomenological quantity which
takes into account the non perturbative long distance physics, including
initial state interactions, and plays, for single spin asymmetries in
$p\,p$ collisions, the same role plaid by the distributions functions
$f_{a/p}(x_a)$ in unpolarized processes. This function is zero
in the absence of parton intrinsic motion, but, for $k_\perp \not= 0$, allows
non zero single spin asymmetries even taking into account only lowest order
perturbative QCD interactions among the constituents. Measurements of the
asymmetries supply information on $I^{a/p}_{+-}$, like measurements of
unpolarized cross-sections supply information on $f_{a/p}$.

In the next Section we introduce a simple parametrization of $I^{a/p}_{+-}$,
based on our knowledge of the distribution functions, and show that it can
reproduce with good accuracy the data on the single spin asymmetry
\beq
A_{N} = \frac{d\sigma^{\pup p \to \pi X} - d\sigma^{\pdown p \to \pi X}}
{d\sigma^{\pup p \to \pi X} + d\sigma^{\pdown p \to \pi X}} =
\frac{d\sigma^{\pup p \to \pi X} - d\sigma^{\pdown p \to \pi X}}
{2 \, d\sigma^{unp}} \,.
\label{main}
\eeq

\vskip 12pt
\nd
{\bf 3 A simple phenomenological model}
\vskip 6pt

Inserting Eqs. (\ref{gendif}) and (\ref{iap}), with the proper kinematical
factors \cite{bla}, into Eq. (\ref{main}) yields
\beq
A_N = {\displaystyle \frac{ \sum_{a,b,c,d}
\int dx_a dx_b \, d^2\bfk_{\perp a}~I^{a/p}_{+-}(x_a, \bfk_{\perp a}) \,
f_{b/p}(x_b) \, [d\hat\sigma/d\hat t(\bfk_{\perp a})]
 \, D_{\pi/c}(z)/z}
{\sum_{a,b,c,d} \int dx_a dx_b ~ f_{a/p}(x_a) \, f_{b/p}(x_b)
(d\hat\sigma/d\hat t) \, D_{\pi/c}(z)/z}} \,,
\label{asyfin}
\eeq
where $d\hat\sigma/d\hat t$ is the unpolarized cross-section for the
elementary constituent process $ab \to cd$; notice that the dependence on
$\bfk_{\perp a}$ has to be kept into account in such a quantity, otherwise
the numerator of Eq. (\ref{asyfin}) would vanish upon integration over
$\bfk_{\perp a}$ due to the fact that $I^{a/p}_{+-}(x_a, \bfk_{\perp a})$
is an odd function of $\bfk_{\perp a}$, see Eq. (\ref{isiv2}). The constituent
momentum fraction $z$ carried by the pion can be expressed in terms of
$x_a$ and $x_b$ by requiring energy-momentum conservation in the elementary
scattering.

All quantities appearing in Eq. (\ref{asyfin}) are either theoretically or
experimentally known, with the exception of the new quantity
$I^{a/p}_{+-}(x_a, \bfk_{\perp a})$, which, in principle, is measurable
via the single spin asymmetry $A_N$. However, in order to
give an estimate of $A_N$ and see if we can obtain reasonable values within
our approach, we parametrize here $I^{a/p}_{+-}$ in a most simple way.

We assume that the dependence of $I^{a/p}_{+-}$ on $\bfk_{\perp a}$
is sharply peaked around an average value $k_{\perp a}^0 =
\langle\bfk^2_{\perp a}\rangle^{1/2}$, value which may depend on $x_a$;
the $x_a$ dependence of $I^{a/p}_{+-}$ which does not originate from the
$k_{\perp a}$ dependence is taken to be of the simple form
\beq
N_a x_a^{\alpha_a}(1-x_a)^{\beta_a} \,,
\label{ix}
\eeq
so that we approximately have
\barr
 & &
\int d^2\bfk_{\perp a}~I^{a/p}_{+-}(x_a, \bfk_{\perp a}) \,
\frac{d\hat\sigma}{d\hat t}(\bfk_{\perp a}) \nonumber \\
& & = \int \!\!\!\!\!\!\!\!\!\!\!
\raisebox{-.5truecm}{$\scriptstyle (\bfk_{\perp a})_x > 0$}
\!\!\!\!\!\!\!\! d^2\mbox{\boldmath $k$}_{\perp a}
I^{a/p}_{+-}(x_a,\bfk_{\perp a})
\!\left[ \frac{d\hat \sigma}{d\hat t}(+\mbox{\boldmath $k$}_{\perp a}) -
\frac{d\hat \sigma}{d\hat t}(-\mbox{\boldmath $k$}_{\perp a}) \right]
\nonumber \\
& & \simeq \frac {k_{\perp a}^0}{M}
N_a x_a^{\alpha_a}(1-x_a)^{\beta_a}
\left[ \frac{d\hat \sigma}{d\hat t}(+k_{\perp a}^0) -
\frac{d\hat \sigma}{d\hat t}(-k_{\perp a}^0) \right] \,,
\label{kint}
\earr
where $M$ is a hadronic mass scale, $M \simeq 1$ GeV/$c$. A numerical estimate
of $k_{\perp a}^0 /M$ can be found in Ref. \cite{rob} and can
be accurately reproduced by the expression
\beq
\frac{k_{\perp a}^0}{M} = 0.47 \> x_a^{0.68}(1-x_a)^{0.48} \,,
\label{kx}
\eeq
which we adopt in our calculations. Such value of $k_{\perp a}^0$ enters
in the computation of [$d\hat\sigma/d\hat t \, (\bfk_{\perp a})
- d\hat\sigma/d\hat t \, (-\bfk_{\perp a})$]; notice that such a quantity
is ${\cal O}(k_\perp / p_T)$ \cite{siv1}, hence the contribution of
Eq. \ref{kint} is a higher twist one.

In order to give numerical estimates of the asymmetry (\ref{asyfin}) we still
need explicit expressions of the unpolarized distribution functions,
$f_{a,b/p}$, and the fragmentation functions, $D_{\pi/c}$. At this stage we
have only considered contributions from $u$ and $d$ quarks inside the
polarized proton, which certainly dominate at large $x_F$ values, that is
$a = u,d$ in the numerator of Eq. (\ref{asyfin}). Instead, we have considered
all possible constituents in the unpolarized protons, with $\bfk_{\perp} =0$,
and all possible constituent fragmentation functions. We have taken
$f_{q,\bar q,g/p}$ from Ref. \cite{mrsa}, $D_{\pi/q,\bar q}$ from
Ref. \cite{field} and $D_{\pi/g}$ from Ref. \cite{cut}. Given the very
limited $p_T$ range of the data we have neglected the QCD $Q^2$ dependence
of the distribution and fragmentation functions.

By using Eqs. (\ref{kint}) and (\ref{kx}) into Eq. (\ref{asyfin}),
together with the unpolarized $f_{a,b/p}$ and $D_{\pi/c}$ functions, we remain
with an expression of $A_N$ still dependent on a set of 6 free parameters,
namely $N_a, \alpha_a$ and $\beta_a$ ($a = u,d$), defined in Eq. (\ref{ix}).
We have obtained a best fit to the data \cite{ada2}, shown in Fig. 1,
with the following values of the parameters:
\beq
\begin{array}{lccc}
\rule[-0.6cm]{0cm}{1.3cm}
 \; & N_a & \alpha_a & \beta_a \\
\rule[-0.6cm]{0cm}{0.5cm}
 u \;\;\;   & 5.19  & 2.79       &  4.15     \\
\rule[-0.6cm]{0cm}{0.5cm}
 d \;\;\;   & \!\!\!\!\! -2.29 & 2.11       &  4.70     \\
\label{param}
\end{array}
\eeq
As the experimental data \cite{ada2} cover a $p_T$ range between 0.7 and
2.0 GeV/$c$ we have computed $A_N$ at a fixed value $p_T = 1.5$ GeV/$c$.
Our asymmetry decreases with increasing $p_T$ and increases at smaller
$p_T$; however, we do not expect our approach to be valid at $p_T$ smaller
than, say, 1 GeV/$c$. We will further comment on the $p_T$ range of our
computation in the conclusions.

Notice that the above values (\ref{param}) are very reasonable indeed;
actually, apart from an overall normalization constant, they might even have
been approximately guessed. The exponents $\alpha_{u,d}$ and $\beta_{u,d}$ are
not far from the very na\"{\i}ve values one can obtain by assuming, as somehow
suggested by Eqs. (\ref{iap}) and (\ref{g}), that  $I^{a/p}_{+-}(x_a)
\sim \sqrt {f_{a,+/p,+}(x_a)f_{a,-/p,+}(x_a)}$, where
$f_{a,+(-)/p,+}(x_a)$ denotes, as usual, the
number density of quarks with the same (opposite) helicity as the parent
proton. Also the relative sign and strength of the normalization constants
$N_u$ and $N_d$ turn out not to be surprising if one assumes that there
might be a correlation between the number of quarks at a fixed value of
$\bfk_{\perp a}$ and their polarization: remember that, according to $SU(6)$,
inside a proton polarized along the $\hat{\bfy}$ direction, $P_y=1$, one has
for valence quarks $P_y^u = 2/3$ and $P_y^d = -1/3$.

It might appear surprising to have approximately opposite values for the
$\pi^+$ and $\pi^-$ asymmetries, as the data indicate, and a large
positive value for the $\pi^0$; one might rather expect $A_N \simeq 0$ for a
$\pi^0$. However, this can easily be understood from Eq. (\ref{asyfin}) which
we simply rewrite, for a pion $\pi^i$, as  $A_N^i = {\cal N}^i/{\cal D}^i$,
if one remembers that from isospin symmetry one has:
\beq
D_{\pi^0/c} = \frac{1}{2}\left( D_{\pi^+/c} + D_{\pi^-/c} \right)\,.
\label{iso}
\eeq

Eq. (\ref{asyfin}) show that the relation
(\ref{iso}) also holds for ${\cal N}^0$ and ${\cal D}^0$, so that
\beq
A^0_N = \frac{{\cal N}^+ + {\cal N}^-}{{\cal D}^+ + {\cal D}^-} =
A^+_N \;\, \frac{ 1 + \frac{\displaystyle A^-_N}{\displaystyle A^+_N}
\frac{\displaystyle {\cal D}^-}{\displaystyle {\cal D}^+}}
{1 + \frac{\displaystyle {\cal D}^-}{\displaystyle {\cal D}^+}} \,\cdot
\label{a0}
\eeq
It is then clear that $A^-_N \simeq - A^+_N$ implies $A^0_N \simeq 0$ only
if $\cal D^- \simeq \cal D^+$, {\it i.e.} if the unpolarized cross-sections
for the production of a $\pi^-$ and a $\pi^+$ are approximately equal.
This is true only at $x_F \simeq 0$. At large $x_F$ the minimum value of
$x_a$ kinematically allowed increases and the dominant contribution to
the production of $\pi^+$ and $\pi^-$ comes respectively from $f_{u/p}(x_a)$
and $f_{d/p}(x_a)$ [see the denominator of Eq. (\ref{asyfin})]. It is
known that $f_{d/p}(x_a)/f_{u/p}(x_a) \to 0$ when $x_a \to 1$; this implies
that ${\cal D}^-/{\cal D}^+$ decreases with increasing $x_F$, so that {\it at
large} $x_F$ we have ${\cal D}^-/{\cal D}^+ \ll 1$ and  $A^0_N \simeq
A^+_N( 1 - 2{\cal D}^-/{\cal D}^+) \simeq  A^+_N$. Such a trend emerges both
from the experimental data and our computations.

\vskip 12pt
\nd
{\bf 4 Conclusions}
\vskip 6pt

We have applied the QCD hard scattering approach, based on the factorization
theorem in the helicity basis, to the description of single spin asymmetries
in inclusive production of pions, with $p_T$ values in the 1 to 2 GeV/$c$
range, in the scattering of polarized protons off unpolarized ones. Such
region of $p_T$ is a delicate one; even if we assume, somewhat optimistically,
that the hard scattering scheme is suitable for the description of these
processes, we expect that, due to the moderate $p_T$ values, higher twist
contributions may still be sizeable and important. Indeed, the leading twist
contribution to the single spin asymmetries is zero and the large experimental
data should be explained via non leading terms.

In evaluating them one has to assume that the factorization theorem still
holds at higher twist level; moreover, one can only take into account a few out
of the many higher twist contributions and corrections to the simple hard
scattering formulae. Here, we have considered the role of the parton
intrinsic $\bfk_\perp$ in the initial polarized proton \cite{siv1, siv2},
while neglecting other possible effects due, {\it e.g.}, to transverse
$\bfk_\perp$ in the fragmentation process. In this our model can only be
regarded as a phenomenological approach to the description of the
otherwise mysterious large single spin asymmetries. Further application
of the same model should test its validity.

We have shown how single spin asymmetries can be different from zero and
originate from lowest order perturbative QCD interactions, provided some
non perturbative and intrinsic $\bfk_{\perp}$ effects are properly taken
into account. In the helicity basis, suitable for the use of the factorization
theorem, this non perturbative long distance information is contained in the
function $I^{a/p}_{+-}(x, \bfk_{\perp})$, which has a simple parton model
interpretation in the transverse spin basis, Eqs. (\ref{isiv1})-(\ref{iap}).

Detailed data on single spin asymmetries in $p\,p$ inclusive interactions
would yield information on $I^{a/p}_{+-}(x, \bfk_{\perp})$, similar
to the information gathered on the unpolarized structure functions; a
knowledge of $I^{a/p}_{+-}$ from some process could then be used to predict
spin effects in other processes. Here we have shown how a simple and realistic
parametrization of $I^{a/p}_{+-}$ can easily explain the data on single spin
asymmetries for the process $p^{\uparrow} p \to \pi X$ at large energy and
$p_T \simeq 2$ GeV/$c$. Our model clearly exhibits the observed increase
with $x_F$, at fixed $p_T$ values, of the magnitude of the asymmetries.
A more detailed application of our approach to other processes, like
$\bar p^{\uparrow} p \to \pi X$, $p^{\uparrow} p \to \gamma X$
and $\pi p^{\uparrow} \to \pi X$ is in progress \cite{noi}: this should
help in assessing the relevance and importance of our estimates.

\vskip 24pt
\nd
{\bf Acknowledgements}
\vskip 6pt
\nd
We would like to thank J.C. Collins for useful correspondence

\vskip 24pt
\noindent
{\bf Figure caption}
\vskip 6pt
\noindent
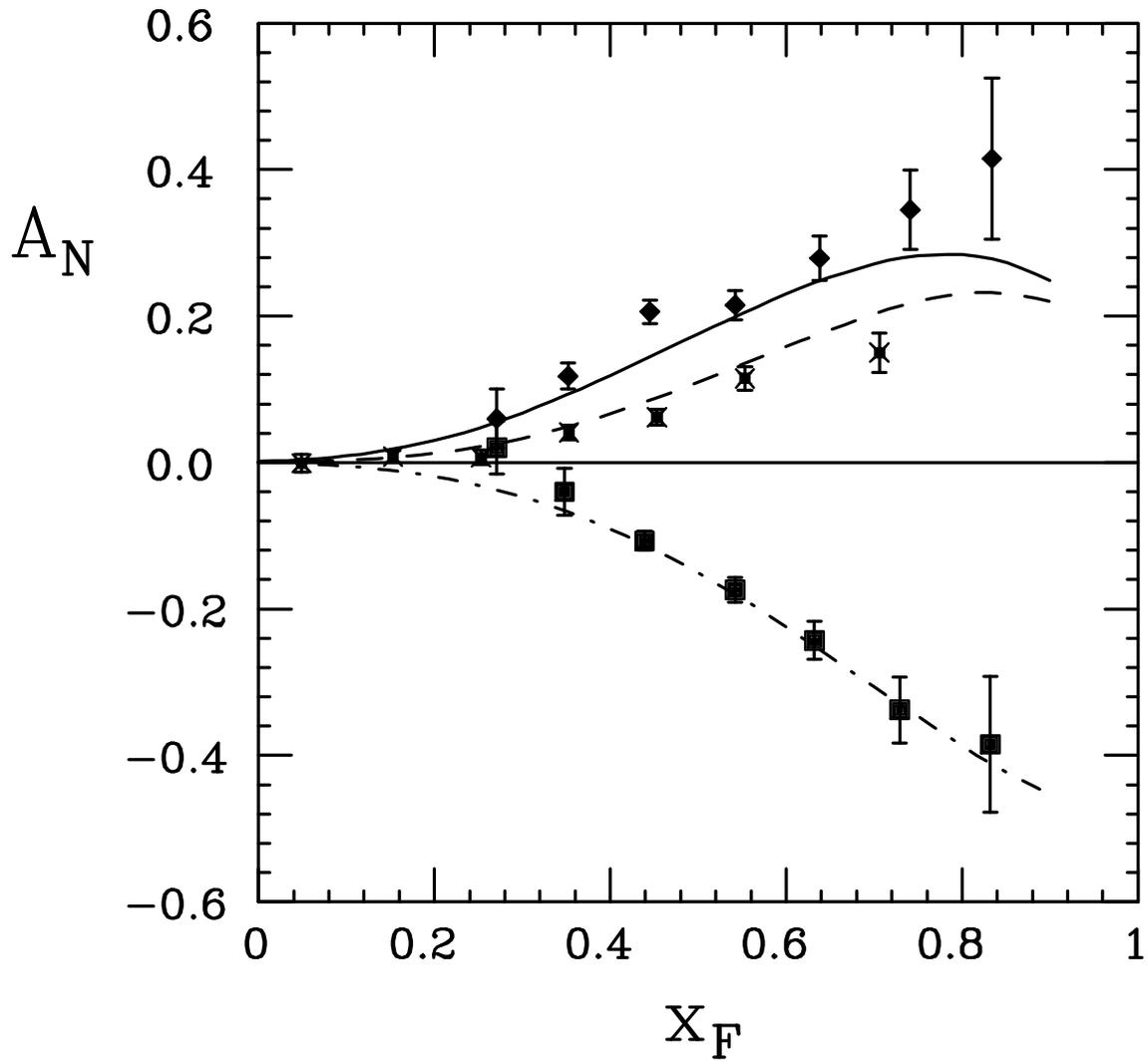
{\bf Fig. 1} Fit of the data on $A_N$ \cite{ada2}, with the parameters
given in Eq. (\ref{param}); the upper, middle and lower sets of data
and curves refer respectively to $\pi^+, \pi^0$ and $\pi^-$.
\end{document}